\documentclass[aps,twocolumn,prd,epsf]{revtex4}
\usepackage{amssymb}
\usepackage{graphicx}
\usepackage{soul}
\usepackage{amsmath}
\usepackage{amsfonts}
\usepackage{comment} 
\usepackage{placeins}
\usepackage{float}
\usepackage{mathrsfs}
\usepackage{lipsum}
\newcommand{\be}{\begin{equation}}
\newcommand{\ee}{\end{equation}}
\newcommand{\bea}{\begin{eqnarray}}
\newcommand{\eea}{\end{eqnarray}}

\newcommand{\fnl}{f_{\rm NL}}

\usepackage{color}

\begin{document}

\title{The primordial curvature perturbation from lattice simulations }
\author{Shailee V. Imrith}
\email{s.v.imrith@qmul.ac.uk}
\author{David J. Mulryne}
\email{d.mulryne@qmul.ac.uk}
\affiliation{\vspace{2mm}
School of Physics and Astronomy, Queen Mary University of London, Mile End Road, London, E1 4NS, UK}
\author{Arttu Rajantie }
\email{a.rajantie@imperial.ac.uk}
\affiliation{\vspace{2mm}
Department of Physics, Imperial College London, London, SW7 2AZ, UK}

\begin{abstract}
We study the contribution to the primordial curvature perturbation on observational scales generated by the reheating field in 
massless preheating. To do so we use lattice simulations and a recent extension to the 
$\delta N$ formalism. The work demonstrates the functionality of these techniques for calculating the observational 
signatures of models in which non-perturbative reheating involves a light scalar field.

\end{abstract}
\maketitle

\section{Introduction}
Inflation is the favoured theory for the production of 
primordial perturbations which seed anisotropies observed in the cosmic microwave 
background (CMB). These perturbations are constrained to be
close to scale invariant and close to Gaussian \cite{Ade:2015ava,Akrami:2018odb}, with tight 
bounds on the spectral index, the tensor to scalar ratio, and the
local non-Gaussianity parameter $f_{\rm nl}^{\rm loc}$. 
It is therefore imperative that we are able to accurately calculate the statistics of perturbations produced by different 
models of the early universe 
 to confront them with observations. For simple models the procedure to do this is well understood, but 
if more than one field is light at the end of inflation, perturbations produced by inflation evolve during the reheating phase which follows. 
This is both a problem and potentially an opportunity, since it makes model predictions sensitive to this complicated and 
poorly constrained phase
of evolution. 
In this paper, for the first time we calculate the statistics of the curvature perturbation on observational scales 
directly using lattice simulations of reheating. 

During reheating, particles may be produced rapidly due to non-perturbative reheating processes, 
often called preheating \cite{Kofman:1997pt} \cite{Kofman:1994rk} \cite{Kofman:1995fi} \cite{Kofman:1997yn}.
If particles are produced in a field that is light during inflation, this process can alter the statistics of the primordial 
curvature perturbation $\zeta$ over observable scales~\cite{Chambers:2007se,Chambers:2008gu}. This is because the rate of particle production is sensitive to 
the initial conditions of the reheating field, which are modulated over such scales. The simplest example is 
massless preheating, in which particles are produced by parametric resonance of a light 
reheating field, and which serves as a testing ground of methods to calculate the statistics of $\zeta$ after inflation. It is the focus of this paper. 

There are two main problems for calculating $\zeta$ after massless preheating.  
First, as a non-perturbative process involving 
the excitation of the inhomogeneous subhorizon modes of the fields, it must be studied using lattice simulations. Second, since these 
simulations typically only capture the dynamics of a patch of the universe exponentially smaller than observable 
universe, and because the average evolution of such a patch can be very sensitive  to 
initial conditions in a highly non-linear way, widely used methods such as the standard $\delta N$ formalism cannot be used without modification. 
These issues have been discussed at length in a number 
of publications  \cite{Chambers:2007se,Chambers:2008gu,Chambers:2009ki,Bond:2009xx}\cite{Kohri:2009ac}\cite{Suyama:2013dqa}.
In our own earlier work on the issue \cite{Imrith:2018uyk}, we discussed the use of a non-perturbative approach to $\delta N$ 
to address these issues, which is 
described below. In this current work, we implement
this method to calculate the statistics of $\zeta$ after massless preheating directly from lattice simulations.

\section{Non-Perturbative $\delta N$ Formalism}
The $\delta N$ formalism \cite{Lyth:2004gb,PhysRevD.42.3936,Sasaki:1995aw,Sasaki:1998ug,Starobinsky:1986fxa} employs the separate universe assumption to calculate the observable 
statistical properties of $\zeta$ produced by scalar field models of the early universe. It typically 
assumes that the number of e-folds, $N$, undergone by 
a patch of the universe coarse-grained on observable scales is a function of the field's initial conditions coarse-grained on those 
scales, and that this function can be
well described by the leading terms in a Taylor expansion. This is used to relate the statistics of field perturbations, 
at some initial time on a flat hypersurface, to the statistics of $\zeta$ at some final time.

In \cite{Imrith:2018uyk}, we discussed how to proceed if 
the evolution of a given patch of the universe is 
highly sensitive to initial conditions and
a Taylor expansion of the $N$ function is not a 
good approximation, as is the case for massless preheating. 
This begins with the definition of the ensemble average of any $n$-point correlation function in real space. 
By working in real space 
there is no
implicit assumption about coarse-graining at the outset and 
the method can therefore utilise simulations of only small regions of the universe 
as long as they do not interact with one another, and hence obey the separate universe assumption. 
In principle the full $n$-point function
can be evaluated directly, but this is often challenging to do. 

Alternatively, one can  
employ 
an expansion which does not assume a Taylor expansion of the e-fold function, but instead a Taylor expansion in the correlation 
between the field's value at separated spatial positions
~\cite{Suyama:2013dqa,Bethke:2013aba,Imrith:2018uyk}. 
We showed \cite{Imrith:2018uyk} that this approach leads to expressions for the correlation functions of 
$\zeta$ which take the same form as those of the usual $\delta N$ formalism 
but with \textit{non-perturbative $\delta N$ coefficients} instead of the usual derivatives of $N$. 
We observed that in many cases our method showed noticeable improvement over standard $\delta N$ expansion,  and 
argued that it is ideally suited to use with lattice simulations, but we also showed that 
even this improved expansion can break down, and that whether this happens needs to be examined on a case by case basis. 
Our method is broadly 
equivalent to the results of Suyama and Yokoyama \cite{Suyama:2013dqa}, where massless preheating was also analysed, but where only an analytical approximation to the results of lattice simulations was used. 

In the present work we therefore implement the non-perturbative $\delta N$ formalism 
together with lattice simulations to directly study massless preheating. 
Since we are primarily interested in the contribution to $\zeta$ from the reheating field, we will assume 
this field is uncorrelated to the inflaton before the start of the lattice simulations, and will therefore employ the non-perturbative expansion 
formalism including only one field. 
In this case the contribution to the two point function of  $\zeta$ from this field, denoted $\chi$, 
at leading and sub-leading  order in the correlation expansion is given explicitly by
\begin{equation}
\langle \zeta({\bf x}_1) \zeta({\bf x}_2) \rangle = {\tilde N_\chi}^2 \Sigma (r_{12}) +\frac{1}{2} {\tilde N_{\chi \chi}}^2 \Sigma (r_{12})^2\,,
\label{npdn} 
\end{equation}
where $\tilde N_\chi$ and $\tilde N_{\chi \chi}$ are our non-perturbative $\delta N$ coefficients 
\begin{eqnarray}
 \tilde N_\chi &=& \frac{\langle \delta \chi N(\chi) \rangle} {\langle \delta \chi^2    \rangle }\,,
 \tilde N_{\chi \chi} =  \frac{ \langle \delta \chi^2 (N(\chi) - \bar N) \rangle     }{ \langle  \delta \chi^2  \rangle^2}\,.
\label{dNexp} 
\end{eqnarray}
In these expressions for the coefficients, the angle brackets denote an ensemble average with $\chi$ drawn from a single variate 
Gaussian distribution with mean value $\bar \chi$ and variance $\Sigma = \langle \delta \chi({\bf x})  \delta \chi({\bf x}) \rangle$,
and $\Sigma(r_{12})$ is the two point function of field perturbations at separated spatial positions, 
$\Sigma(r_{12}) =  \langle \delta \chi({\bf x_1})  \delta \chi({\bf x_2}) \rangle$. This is evaluated at the initial time before 
reheating, while the two point function of $\zeta$ is evaluated at the final time after reheating. 

We use the sub-leading term above as a necessary test of the validity of the expansion, and assume that only 
in cases where the sub-leading term is sub-dominant 
can we trust our expansion method to calculate $\mathcal{P}_{\mathrm {pre}}$ with reasonable 
accuracy.

We also calculate the local non-Gaussianity using the leading order expression 
for the three point function of $\zeta$ in the correlation expansion
\be
\langle \zeta({\bf x}_1) \zeta({\bf x}_2)  \zeta({\bf x}_3) \rangle = {\tilde N_\chi}^2\tilde N_{\chi \chi} \Sigma (r_{12})\Sigma (r_{23}) + {\rm cyclic}\,.
\ee

\section{Massless Preheating}
Massless preheating \cite{Moghaddam:2014ksa,Greene:1997fu,Prokopec:1996rr}  is defined by the potential
\begin{equation}
V= \frac{1}{4} \lambda \phi^4 + \frac{1}{2} {g}^2 \phi^2 \chi^2,
\label{masslesspreheatingpotential} 
\end{equation}
where $\phi$ is the inflaton field and $\chi$ the reheating field, 
with dimensionless coupling constants $g$ and $\lambda$. 
The first term in Eq.~(\ref{masslesspreheatingpotential}) drives inflation while the second term is 
responsible for particle creation by parametric resonance after inflation ends, 
when oscillations of the inflaton lead to an oscillating mass for the $\chi$ field.
For small coupling ratios , $g^2/\lambda$, required to 
excite the first resonance band of the system,
the masses of the two fields,  $m_{\phi} = \sqrt{3 \lambda}\phi$ and $m_{\chi} = {g} \phi$, 
are comparable during inflation, and hence the  $\chi$ field is light.

This model is unlikely to be compatible with current observations, but 
still serves as a convenient testing ground for our methods. It is conformally invariant 
meaning that the expansion of the universe can be 
re-scaled away simplifying  lattice  simulations. 
During inflation, $\chi$ is approximately zero and the behaviour of the model is the same as the 
standard single field $ \frac{\lambda}{4}\phi^4$ chaotic inflation model. We can therefore estimate 
the value of the $\phi$ field when 
inflation ends by assuming only $\phi$ drives inflation 
and setting the first slow-roll parameter $\epsilon_v = \frac{1}{2}{M_{\mathrm{pl}}^2} (  V_{, \phi }/{V} )^2$ to 1.
We obtain $\phi_{\mathrm{end}} \approx 2.83 M_{\mathrm{pl}}$.

 \subsection{Expressions for $\mathcal{P}_{\mathrm{inf}} $ and $\mathcal{P}_{\mathrm{pre}}$ }

We now begin the process of calculating the impact of reheating on $\zeta$ using the formalism 
outlined above. First let us note that we can immediately estimate the contribution to $\zeta$ from the inflaton, $\phi$, 
under our assumption that it is uncorrelated to the reheating field.
Assuming the observational pivot scale corresponds to modes which crossed the horizon $55$ e-folds before the end of 
inflation, the field value corresponding to this scale is $\phi_{\star} \approx 21.17 M_{\mathrm{pl}}$ .
This value can then be employed in the standard expression for the power spectrum of $\zeta$ 
 \begin{equation}
\mathcal{P}_{\mathrm{inf}} = \frac{1}{8 \pi^2} \frac{H_{\star}^2}{ M^2_{\mathrm{pl}} \epsilon_{v_{\star}} }\approx \frac{\lambda \phi_{\star}^6}{768 \pi^2  M^6_{\mathrm{pl}}   } .
\label{4} 
\end{equation}

On the other hand, to calculate the contribution of the $\chi$ field we must employ the 
non-perturbative $\delta N$ formalism.
The effective mass of the $\chi$ field is $g \phi$, and we must begin the 
$\delta N$ formalism before 
this field becomes heavy. We therefore set initial conditions for $\chi$ and its statistics at the time just before 
$\chi$ becomes massive, i.e., when ${g^2 \phi_{\mathrm{ini}} ^2}/{H_{\mathrm{ini}}^2} \sim 1$. This is 
shortly before the end of inflation. 
At this point we assume again that $\chi$ can be taken to be a spectator field uncoupled to the inflaton, and therefore 
has a variance within our observable universe about some background value, $\bar \chi$, of approximately 
\begin{equation}
\langle \delta \chi \delta \chi \rangle \vert_{\mathrm{ini}} = \frac{H^2_{\mathrm{ini}}}  {4 \pi^2} \mathcal{N}
\label{variance} 
\end{equation}
where  $\mathcal{N} \approx55$ and $H_{\mathrm{ini}}$ is the Hubble rate at this time. 
One then finds the amplitude of  power spectrum of $\zeta$ from preheating 
\begin{equation}
\mathcal{P}_{\mathrm{pre}} = \tilde N_\chi^2 \frac{H_{\mathrm{ini}}^2 } {4 \pi^2}\,.
\label{5} 
\end{equation}
We recall that if our expansion is a good approximation, the $k$ dependence of the spectrum 
is inherited from the spectrum of $\chi$ fluctuations at the initial time, which we take to be close to scale invariant. 
The value of $\bar \chi$ is determined by the behaviour of the system prior to the observable number 
of e-folds, and for the purpose of our study we take it to be a free parameter, with the restriction that 
it must be much greater than the variance of $\chi$ for consistency. 
To evaluate $\tilde N_\chi$ and $\tilde N_{\chi \chi}$ using Eq.~(\ref{dNexp}) requires 
the use of lattice simulations which we now discuss.

\subsection{Simulations}
\label{simulation}   

The $\delta N$ formalism requires that we record the number of e-folds that occur between given initial 
conditions and  some final value for the density of the universe $\rho_{\rm end}$ 
after reheating has occurred and the system has become adiabatic. 
In the case of preheating, this has to be done using lattice field theory simulations. This has been done for massless preheating in Refs.~\cite{Chambers:2007se,Chambers:2008gu,Bond:2009xx}.

In this study, in contrast with previous works, we 
evaluate  $\tilde N_\chi$ and $\tilde N_{\chi \chi}$  directly, 
rather than using some pre-generated or approximated $N(\chi)$ function.
To do so we use the HLattice code \cite{Huang:2011gf} to
simulate small patches of the universe (of order the horizon size at the end of inflation)
from 
the end of inflation until after preheating, and employ a Monte Carlo approach to calculate 
$\tilde N_\chi$ and $\tilde N_{\chi \chi}$. We  describe this process in more detail below. 
We choose HLattice because its symplectic integrator allows for an extremely accurate 
calculation of the number of e-folds. It has previously been used to calculate the change in e-folds due to reheating in
Ref.~\cite{Bond:2009xx}. 

\subsubsection{Monte Carlo Method}

After fixing the parameters of the massless preheating potential, we calculate the energy
scale and the value of $\phi$ at the time at which $\chi$ becomes massive. 
At this time the value of the $\chi$ field is drawn from a Gaussian distribution with mean value $\bar \chi$, and 
variance given by Eq.~(\ref{variance}). As discussed above the mean value, $\bar \chi$,  
can be treated as a free parameter provided 
it is much greater than the variance. 
Using these values of $\phi$ and $\chi$, HLattice then evolves the 
background scalar field equations until the end of inflation before beginning a lattice simulation. 
For the lattice simulation, HLattice treats the initial homogeneous field values as external inputs and
takes the average value of the fields in the lattice patch to be given by the background scalar field values 
at the end of inflation. The fields are also given an inhomogeneous 
(over the lattice) random Gaussian vacuum fluctuations which average to zero.

We modify Hlattice to stop 
the simulation once some final value of the density $\rho = \rho_{\mathrm {end}}$ has been passed. 
We choose a value more than $5$ e-folds after inflation ends. Finally, we use a fitting function to 
calculate $\log a$ as a function of $\log \rho$ over a short range around $\rho_{\mathrm {end}}$ and use this fit to 
calculate an accurate value for $N$ at  $\rho_{\mathrm {end}}$. 
One issue 
we find, as was also reported in Bond et al. \cite{Bond:2009xx}, is that since reheating is not fully complete after preheating, 
the equation of state $\omega = {p}/{\rho}$ oscillates around the radiation dominated value of $1/3$ after preheating, indicating 
that the system is not yet fully radiation dominated and adiabatic.
If the system were to be adiabatic, the difference in $N$  between two different patches with different initial 
conditions and measured at the same 
successive values of $\rho$ would become a constant (i.e., $\zeta$ would become conserved). Instead we find that this 
quantity is oscillating. In order to remedy this problem we follow \cite{Bond:2009xx} and average the result over 
many oscillations. This is then taken to be our 
$N_{\rm end}$, and we 
use this value of $N$ to evaluate the quantity inside the angle brackets on the right hand side of Eq.~(\ref{dNexp}).
By repeatedly 
drawing new values of $\chi$, and repeating the procedure, 
we can calculate the averages in Eq.~(\ref{dNexp}). 
The answer should converge 
as the number of draws increases. We find this does indeed occur with  ${\cal O}(10^4)$
draws needed, and we use this number to generate our results. 
When we come to present our results we further use resampling to estimate their error.
Due to the large number of runs required, we 
 run our simulations on Queen Mary's Apocrita HPC facility \cite{king_thomas_2017_438045}.
For our simulations, we choose a resolution of $32^3$ with comoving box size $20/H$, and choose the LATTICEEASY discretization 
scheme within HLattice, which is suitable because we are ignoring metric perturbations (for more details, see \cite{Huang:2011gf}).

\subsubsection{Checks}
\begin{figure}[]
\centerline{\includegraphics[angle=0,width=80mm, height=51mm]{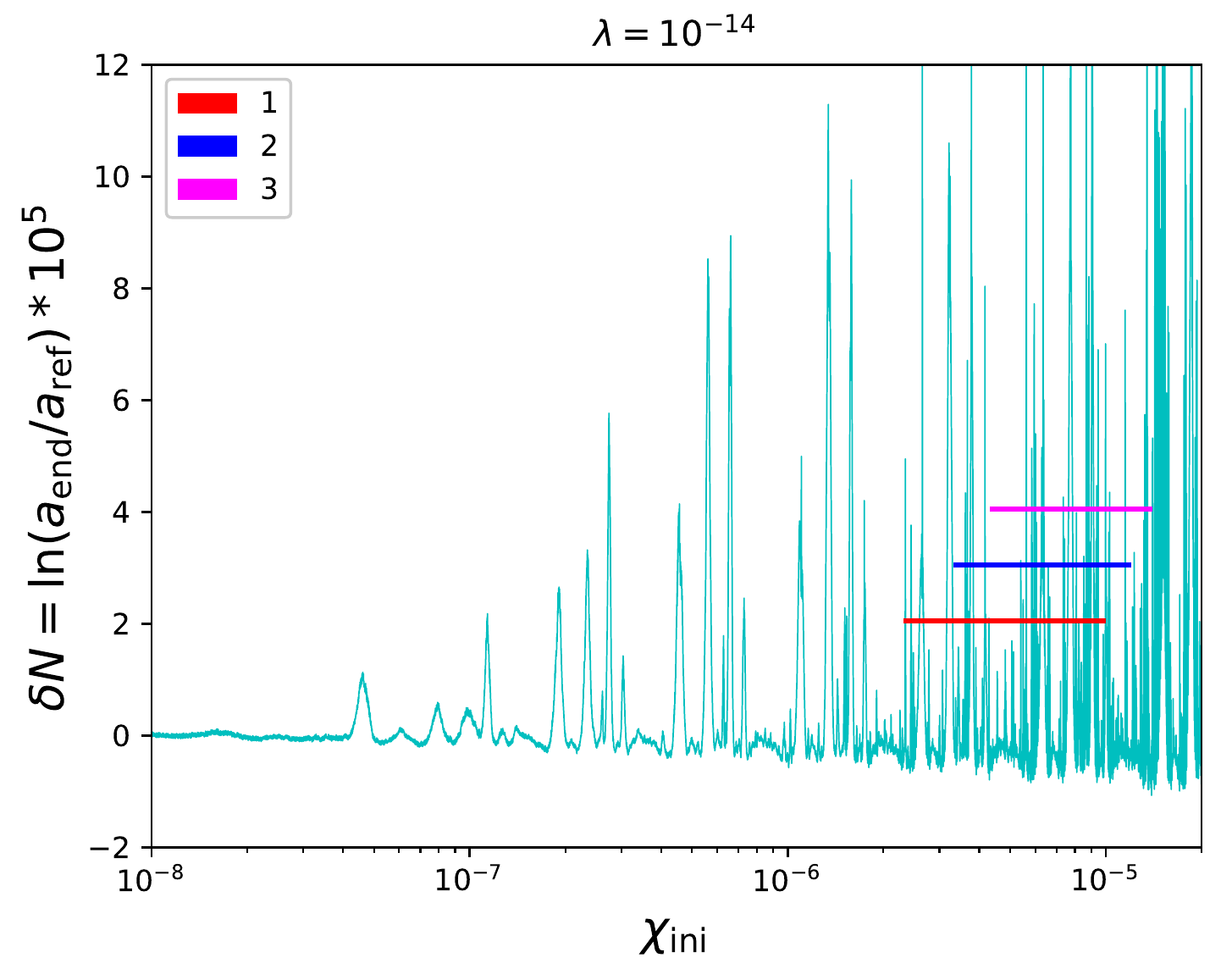}}  
\caption{
The value of $\delta N$ for $\lambda=10^{-14}$, and for different initial field values, $\chi_{\rm ini}$, in Planck units. The three coloured lines (red, blue and magenta) span the range of values that $\chi_{\mathrm {ini}}$ takes, i.e., $\bar \chi_i \pm 3 \sigma$ for $\bar \chi_1=$ $5 \times 10^{-6}M_{\mathrm{pl}}$, $\bar \chi_2 =6 \times 10^{-6}M_{\mathrm{pl}}$ and $\bar \chi_3=7 \times 10^{-6}M_{\mathrm{pl}}$. Note that here, our $\chi_{\mathrm {ini}}$ is the value the field takes just before it becomes massive.}
\label{L14ini}
\vspace*{-15pt}
\end{figure}

To verify the code is working as intended before generating results for our study, 
we first reproduced the plot given in \cite{Bond:2009xx}, where $N_{\rm end}$ was plotted against $\chi$
fixed at the end of inflation (not at a time just before $\chi$ becomes 
massive like we will consider for our purposes --  a similar plot for the initial conditions we use can be seen in 
Fig.~\ref{L14ini}).
The plot is highly featured with many spikes generated by the chaotic behaviour of the system, and 
shows clearly the necessity of using  the non-perturbative $\delta N$ method. 
Moreover, we check that our results are insensitive to the period over which we average the oscillations 
caused by the oscillating equation of state at the end of pre-heating. Finally, 
we also check that $N$ is effectively unchanged when $\chi_{\mathrm{ini}}$ is varied for values of the coupling ratio $g^2/\lambda$ for which the homogeneous 
mode is outside a resonance band of the system. We do this for
$g^2/\lambda =1$ and $3$.

\subsection{Results}

We now present some results for a number of different parameter choices, presenting both the inflationary contribution 
and the contribution from preheating under the assumptions made above.

\subsubsection{Case 1 : $\lambda = 1.8\times 10^{-13}$}
 
We begin by taking $\lambda = 1.8\times 10^{-13}$, with $g^2/\lambda = 2$. 
This value of $\lambda$ leads to a contribution to the power spectrum of
$\mathcal{P}_{\mathrm {inf}} = 2.10 \times 10^{-9}$ from the inflaton field,
in agreement with observations. The model is still not a realistic theory of inflation because the predicted tensor-to-scalar ratio is too high to be compatible with observations~\cite{Akrami:2018odb}.
There may be ways of curing this, for example by having a non-minimal coupling between the inflaton and spacetime curvature, but we will not attempt to do that.

Recalling the discussion below Eq.~(\ref{4}) and using $H^2_{\rm ini} \approx \lambda \phi^4_{\rm ini}/(12 M_{\mathrm{pl}}^2) \approx g^2\phi^4_{\rm ini}/(24 M_{\mathrm{pl}}^2 )$, 
implies that $ H_{\mathrm{ini}} = 3.22 \times 10^{-6} M_{\mathrm{pl}}$. From 
Eq.~(\ref{variance}), we then find $ \langle \delta \chi^2 \rangle = 1.44 \times 10^{-11} M^2_{\mathrm{pl}} $, 
which provides the square of the 
standard deviation of the Gaussian distribution from which we draw the initial 
values of $\chi$.

For the mean of the distribution, $\bar\chi$,
we choose $\bar \chi_i = (i) \times 10^{-5} M_{\mathrm{pl}}$ where $i$ runs from $2$ to $9$,
and run $10^4$ simulations 
for each. Since the data drawn for these values overlap, we are also able to reweight the 
data to calculate results for mean values in between the initial choices. Finally, by resampling the data 
using the so-called 
bootstrap approach, we can determine the approximate one sigma error on our results. 

\begin{figure}[]
\centerline{
\includegraphics[angle=0,width=44mm, height = 35mm]{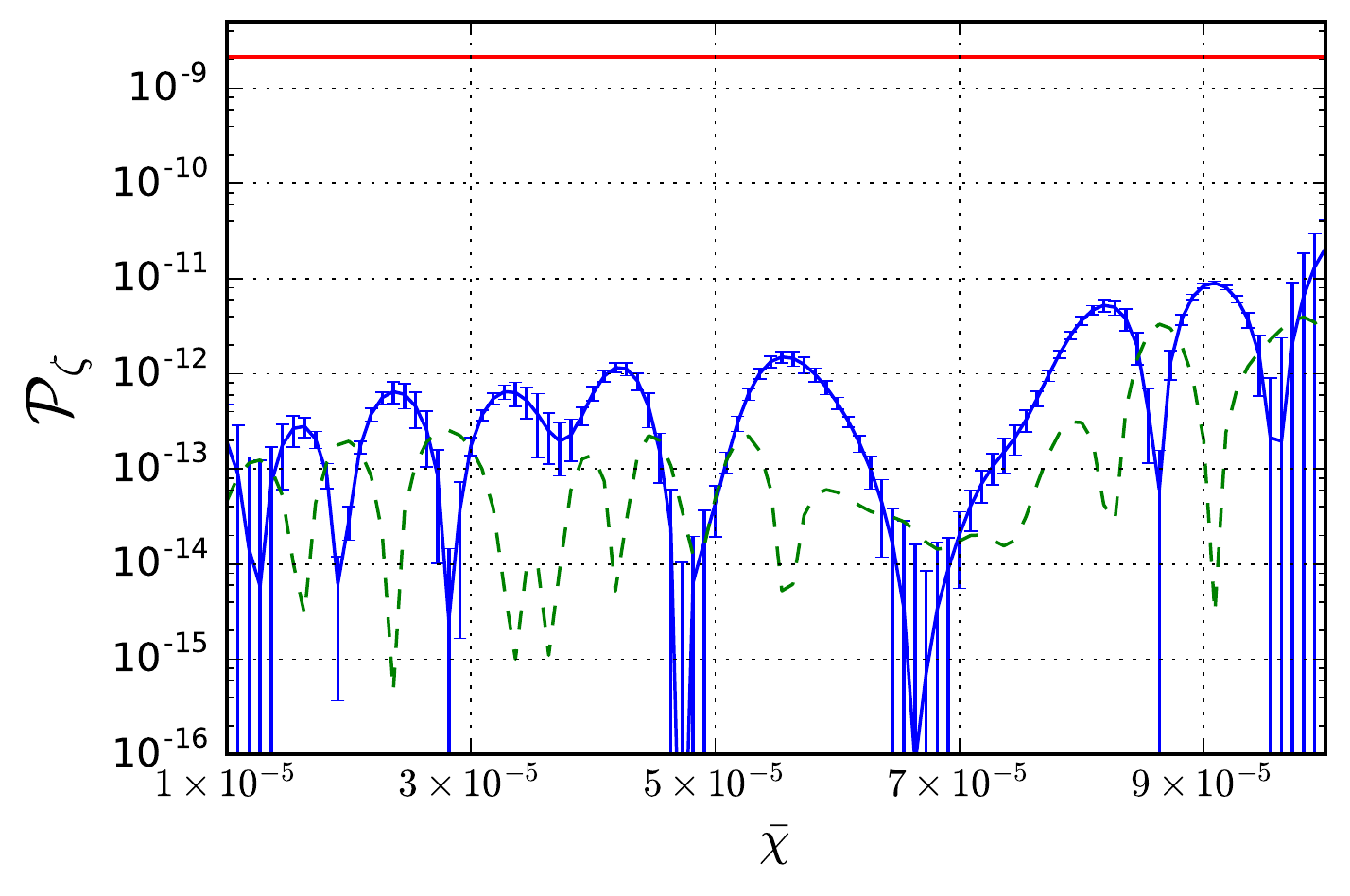}
\includegraphics[angle=0,width=44mm, height = 35mm]{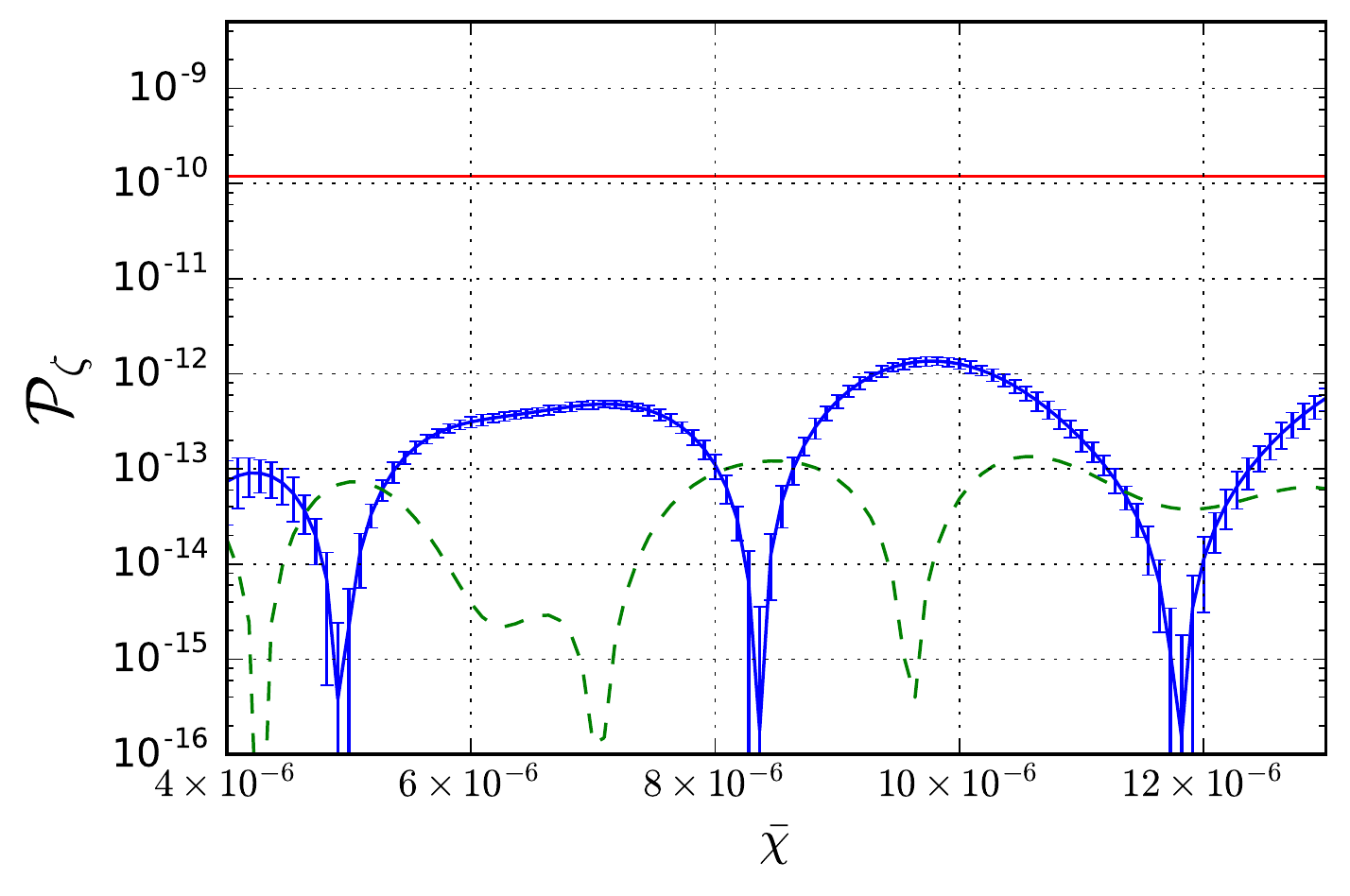} 
} 
\centerline{
\includegraphics[angle=0,width=44mm, height = 35mm]{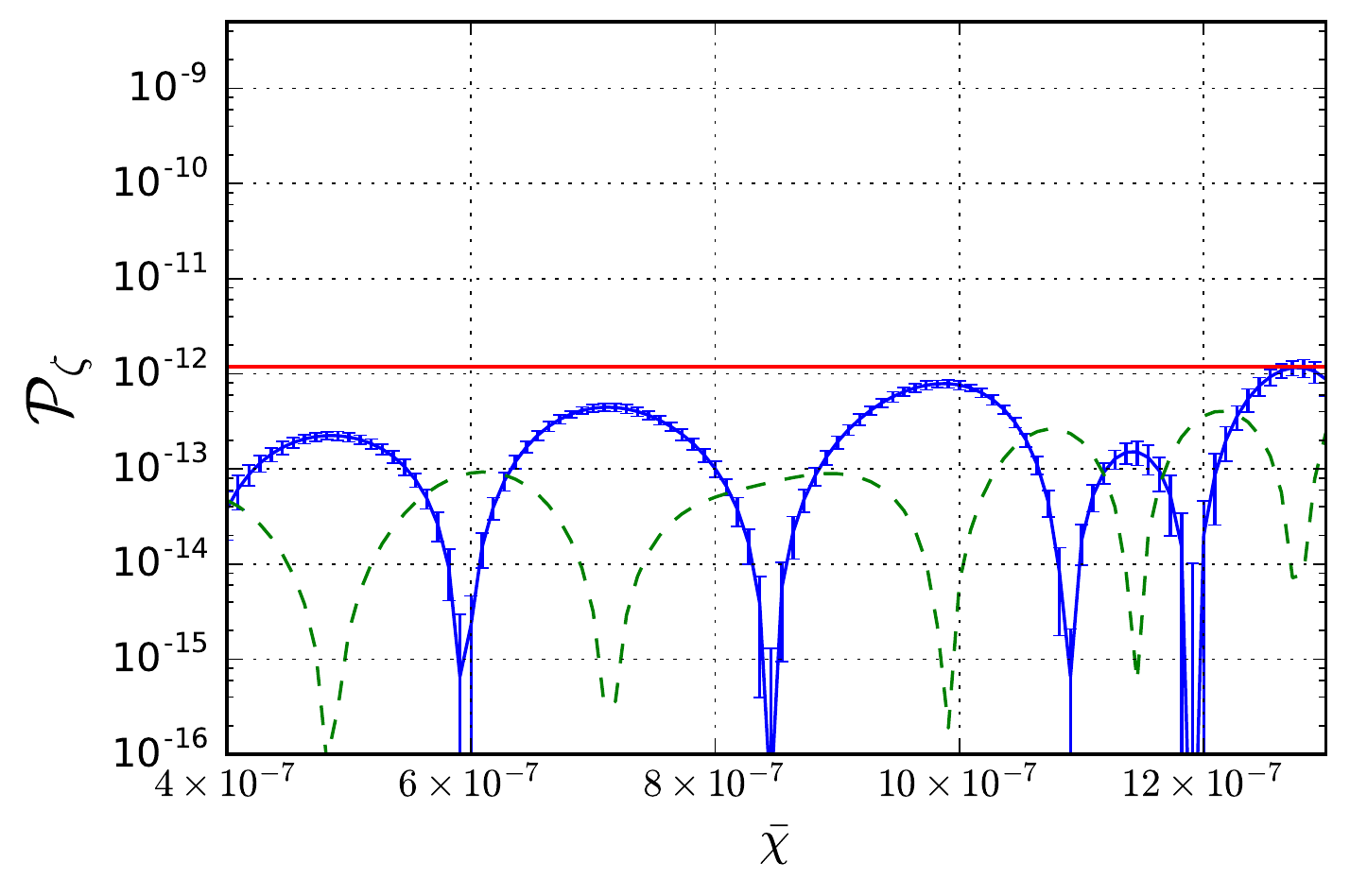}
\includegraphics[angle=0,width=44mm, height = 35mm]{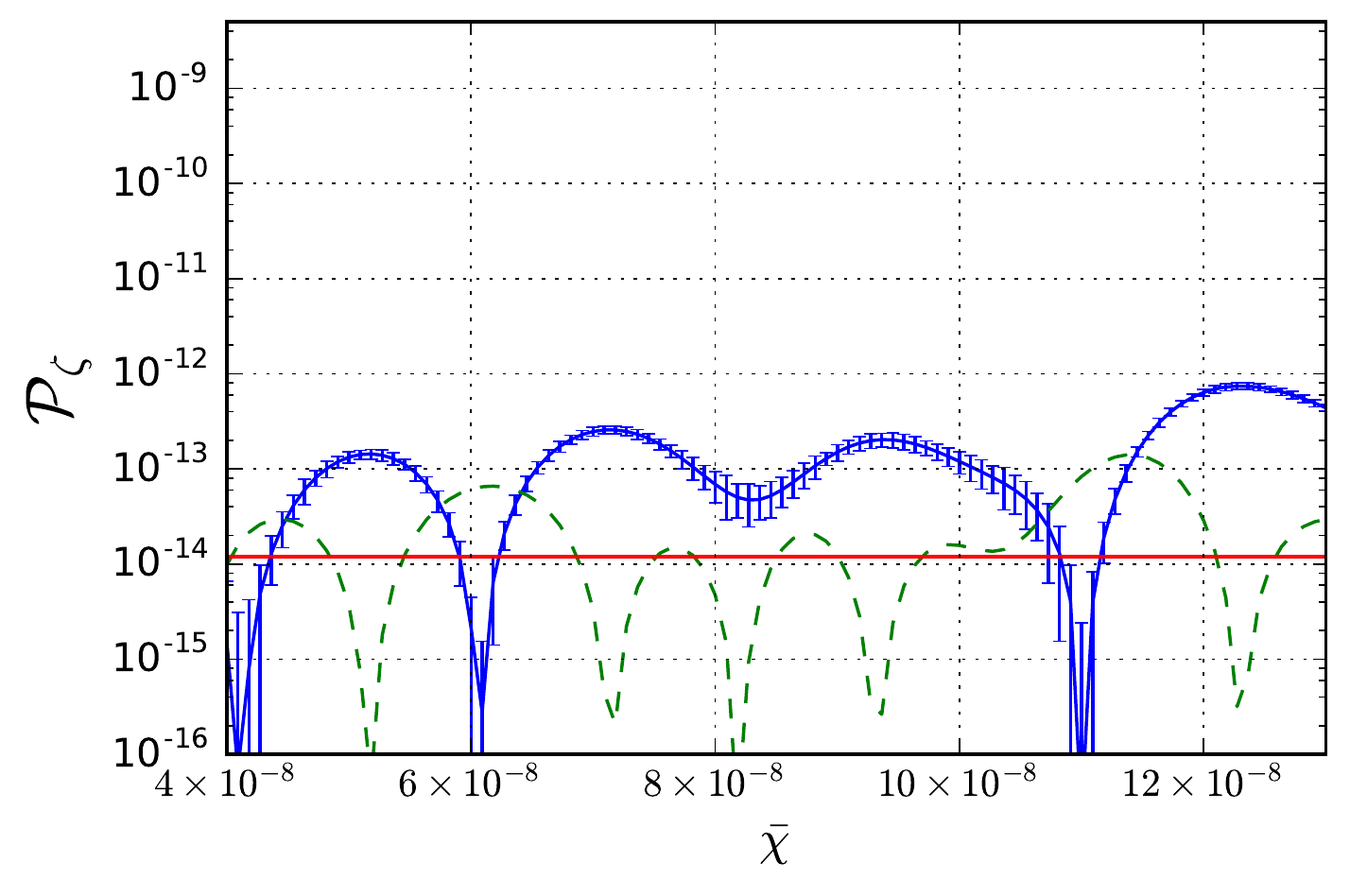} 
} 
\caption{The power spectrum of curvature perturbations produced by preheating, ${\cal P}_{\rm pre}$, as a function of $\bar{\chi}$ in Planck units. Four cases cases are presented. 
In Case 1, top left, $\lambda = 1.8\times10^{-13}$; 
in Case 2, top right,  $\lambda = 10^{-14}$;  
in Case 3, bottom left,  $\lambda = 10^{-16}$; 
and Case 4, bottom right, $\lambda = 10^{-18 }$.
The horizontal red lines show the contribution ${\cal P}_{\rm inf}$ from the inflaton field, and the dashed lines show the sub-leading contribution from Eq.~(\ref{npdn})}.
\label{Pplot}
\vspace*{-15pt}
\end{figure}

The results together with uncertainties are shown in Fig.~\ref{Pplot}. The horizontal 
line shows the amplitude $\mathcal{P}_{\mathrm {inf}}$ of the perturbations generated by the inflaton field.
In Fig.~\ref{L14ini} for illustrative purposes we also show 
how the $N(\chi)$ function looks for $\lambda=10^{-14}$. The three coloured lines superimposed on the 
plot are centred on three of our choices for $\bar \chi$, and indicate the  $3\sigma$ 
range around this value. Fig.~\ref{Pplot} indicates the results are highly sensitive to the initial $\bar \chi$, and 
this can be understood by considering Fig.~\ref{L14ini} which shows that the form of the $N$ functions sampled 
when drawing $\chi$ from a Gaussian distribution changes dramatically depending on the value of $\bar \chi$.

As we have described, the method we have outlined relies on an expansion in the probability distribution 
for field values at separated positions in real space. The criterion we use for the validity of the 
expansion requires 
that the leading term in Eq.~(\ref{npdn}) is dominant over the sub-leading one.
This criterion is  scale dependent and gets worse for shorter 
scales. Here, we consider the criterion for scales which exited $5$ e-folds after the largest observable scale exits the 
horizon, which roughly corresponds to the shortest scale observed on the CMB. The power spectrum 
resulting from the sub-leading term on this scale is given by the dashed line in Fig.~\ref{Pplot} (see for example 
Ref.~\cite{Lyth:2006gd} for a calculation of the power spectrum associated with this term). We find that the expansion is valid for most of the parameter range we 
study, but breaks down in the regions on the plot where the magnitude of the power spectrum  
as calculated using the leading term drops towards zero. These 
regions are also accompanied by large looking error bars on our logarithmic plots. In these regions, producing 
more accurate results would
require the fully non-perturbative formulae for $\mathcal{P}_\zeta$, as described (and performed) in 
Ref.~\cite{Imrith:2018uyk}.

\begin{figure}[]
\centerline{
\includegraphics[angle=0,width=44mm, height = 35mm]{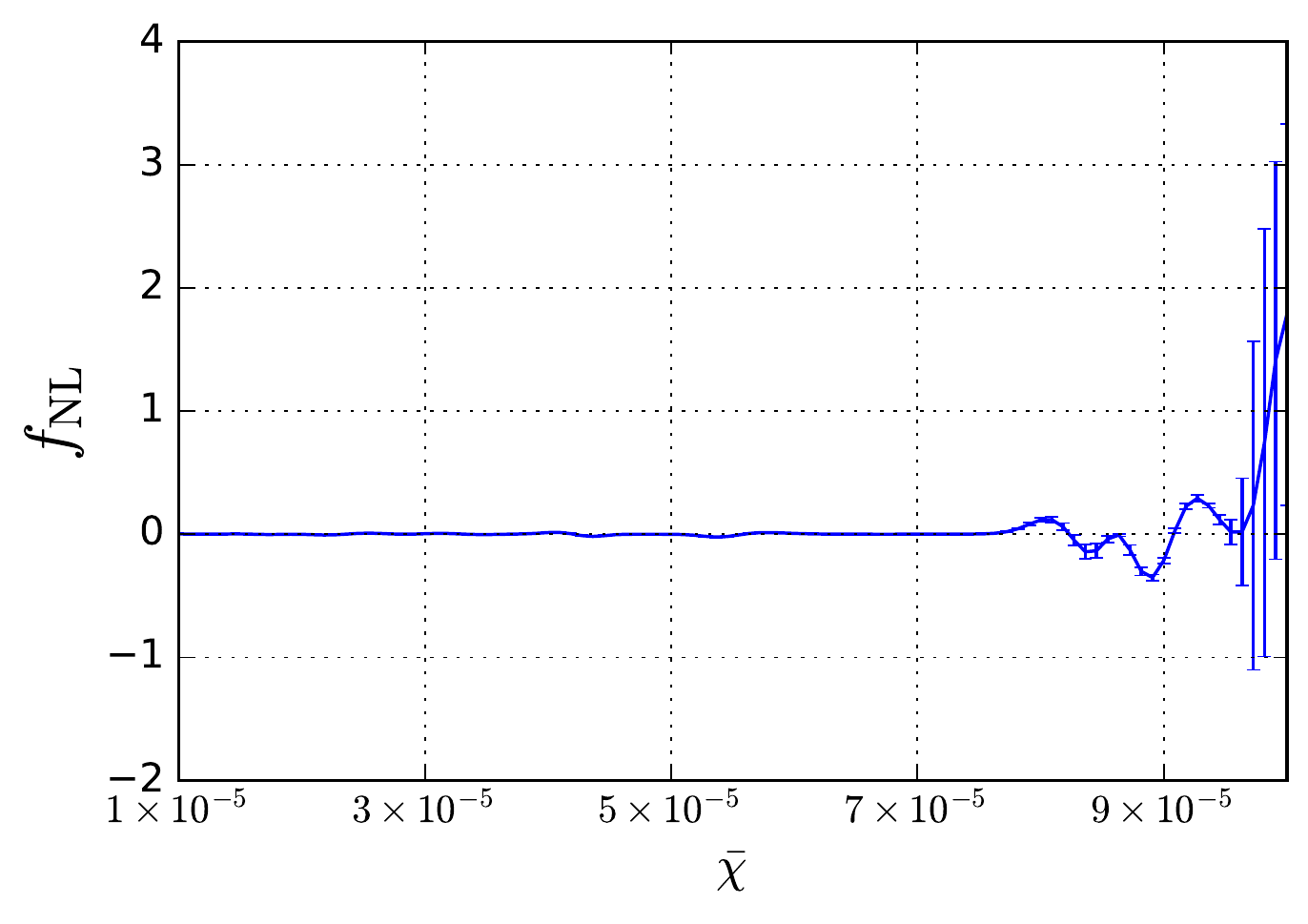}
\includegraphics[angle=0,width=44mm, height = 35mm]{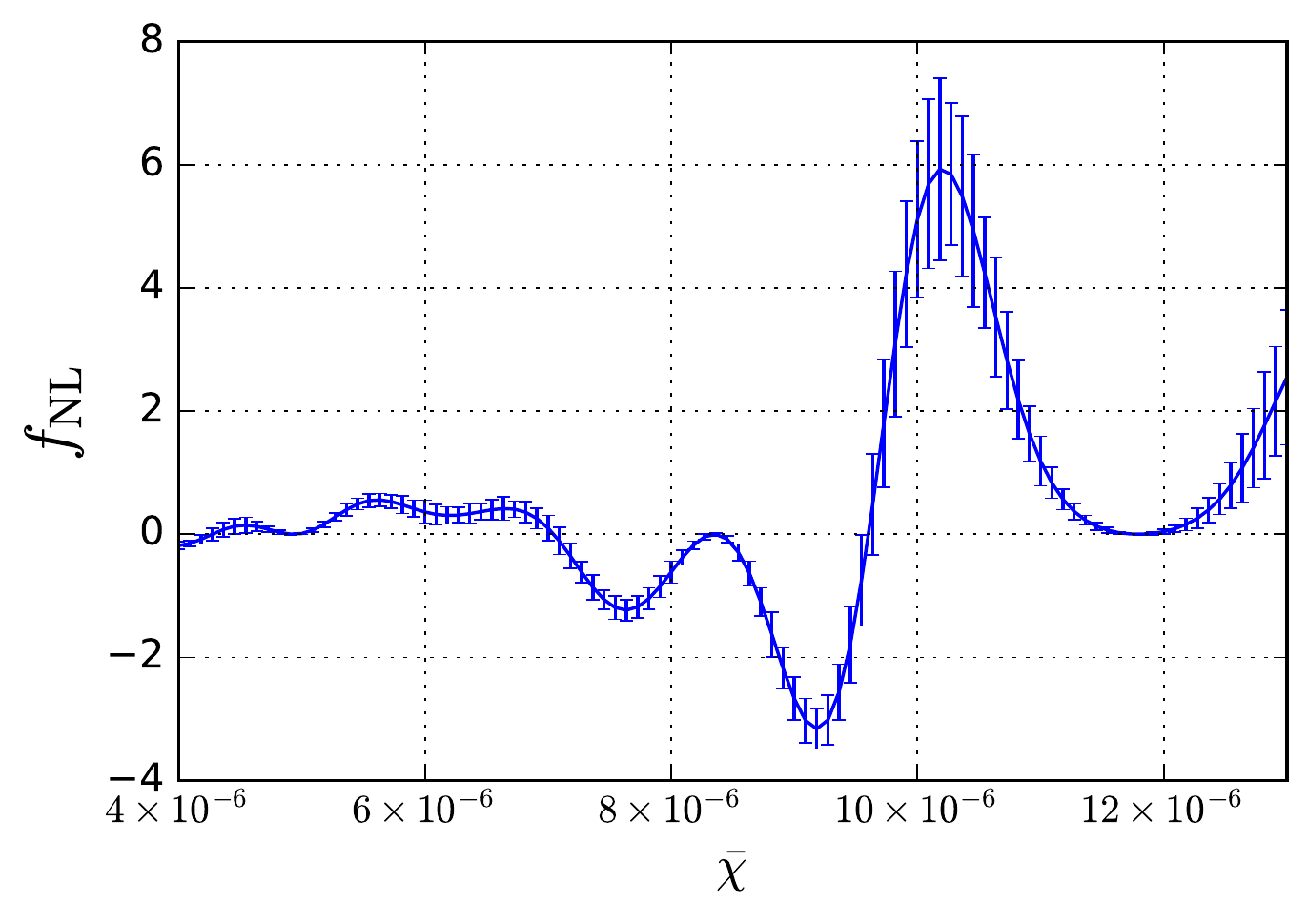} 
} 
\centerline{
\includegraphics[angle=0,width=44mm, height = 35mm]{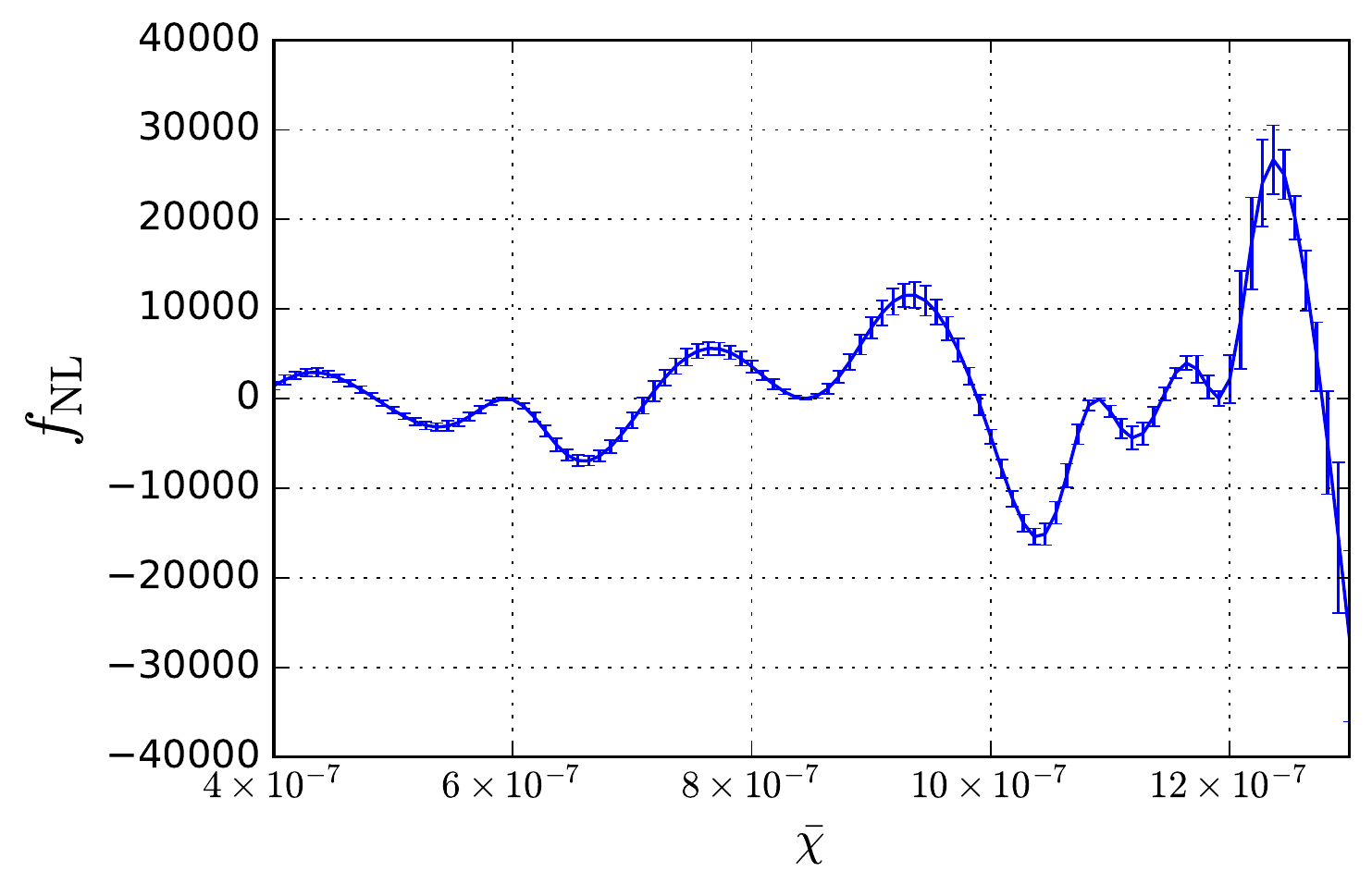}
\includegraphics[angle=0,width=44mm, height = 35mm]{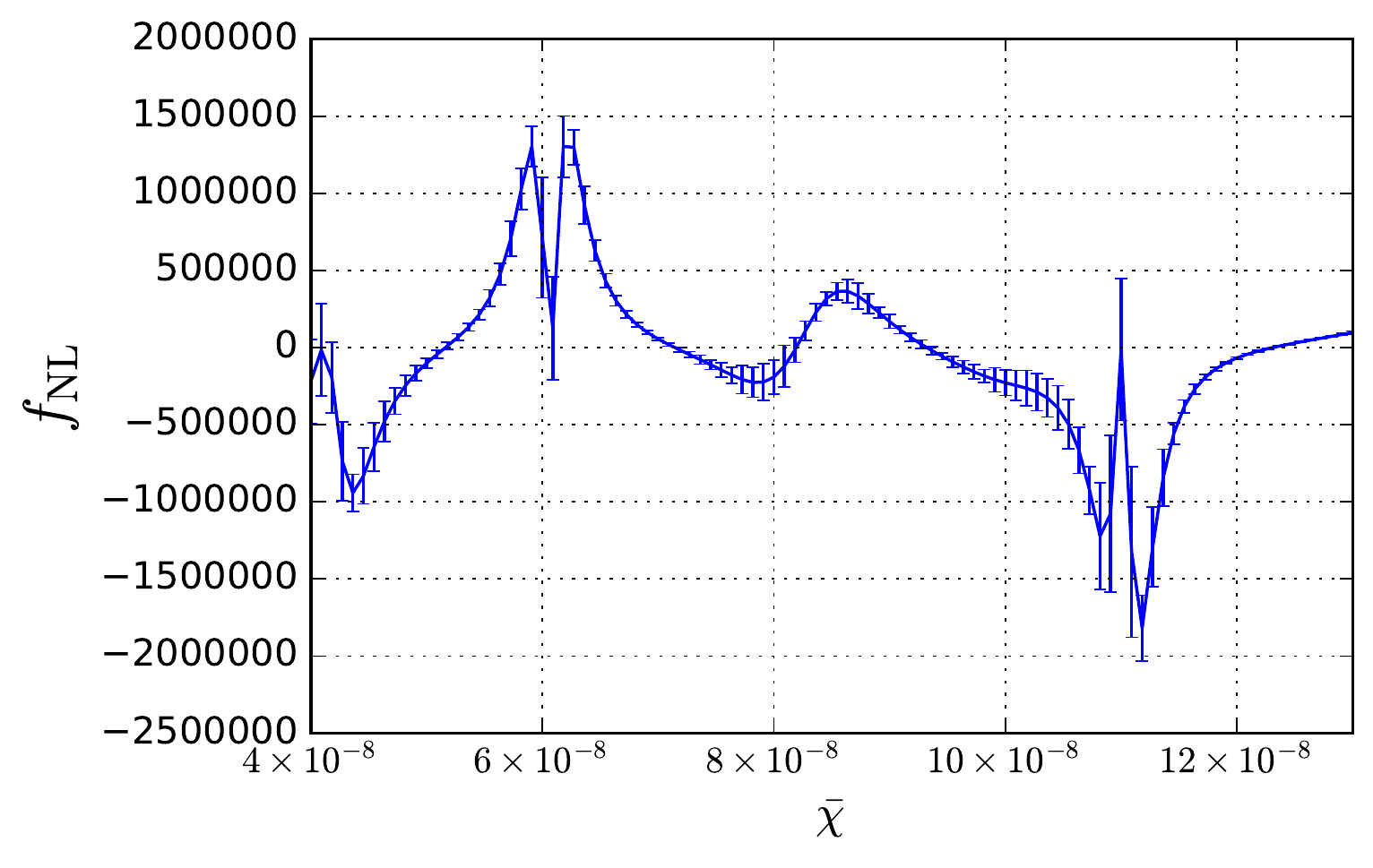} 
} 
\caption{The non-Gaussianity parameter $\fnl$ as a function of $\bar{\chi}$ in Planck units. Four cases cases are presented. 
In Case 1, top left, $\lambda = 1.8\times10^{-13}$; 
in Case 2, top right,  $\lambda = 10^{-14}$;  
in Case 3, bottom left,  $\lambda = 10^{-16}$; 
and Case 4, bottom right, $\lambda = 10^{-18 }$.}
\label{fnlplot}
\vspace*{-15pt}
\end{figure}

The contribution to the curvature perturbation $\zeta$ from preheating is  sub-dominant compared with the inflaton 
contribution ${\cal P}_{\rm inf}$, and therefore not observable in the spectrum. 
This is similar to the results of  \cite{Suyama:2013dqa}, though we find that the 
preheating field's contribution is at least an order of magnitude larger than that calculated there --  reinforcing 
the importance of working directly with lattice simulations.
However, because the inflaton contribution is highly Gaussian, it is
interesting to see whether the preheating contribution could be observed through its non-Gaussianity.
To do this, we calculate the conventional non-Gaussianity parameter $\fnl$
from the expression 
\begin{equation}
 f_{\mathrm {NL}} \approx \frac{\tilde N_{\chi \chi} \tilde N_{\chi} \tilde N_{\chi} H^4_{\rm ini} }{( \frac{1}{2 \epsilon*} \frac{H_*^2}{M^2_{\rm {pl}}} + \tilde N_{\chi} \tilde N_{\chi} H^2_{\rm ini}      )^2 }\times  \frac{5}{6}\,,
\label{observedfnl} 
\end{equation}
where we have assumed that field fluctuations before reheating are Gaussian, and 
the inflationary contribution to $\zeta$ is Gaussian. The results are presented in Fig.~\ref{fnlplot}. We can see that although $\fnl$ is generally small, it becomes significantly larger than 
typical single field value for certain values of $\bar\chi$, making it potentially observable.
This suggests
that it may also be possible to achieve an observable contribution to $\fnl$ in more realistic models.

\subsubsection{Case 2 : $\lambda = 10^{-14}$}

As an alternative to Case 1, 
we also 
consider cases in which the inflaton contribution is 
below the observational 
value amplitude. We calculate the contribution 
from the preheating field to see whether it can be larger than the inflaton contribution,
and whether it could account for the observed perturbation spectrum.
 
We choose $\lambda = 10^{-14}$, again with $g^2/\lambda = 2$. 
This value of $\lambda$ leads to an inflaton contribution to the power spectrum of
$\mathcal{P}_{\mathrm {inf}} =1.19 \times 10^{-10}$, and
Hubble rate $ H_{\mathrm{ini}} = 7.59 \times 10^{-7} M_{\mathrm{pl}}$. The $\chi$ field
variance is $ \langle \delta \chi^2 \rangle = 8.028 \times 10^{-13} M^2_{\mathrm{pl}} $.
For the mean value $\bar\chi$, we choose 
$\bar \chi_i = (i) \times 10^{-6} M_{\mathrm{pl}}$ where $i$ runs from $5$ to $12$, 
and proceed as in Case 1.

As shown in Fig.~\ref{Pplot}, the preheating contribution to $\zeta$ is 
sub-dominant to that produced 
by the inflaton field, and therefore this case does not produce sufficient curvature perturbations to be compatible with observations. The non-Gaussianity parameter $\fnl$ is larger than in Case 1, as shown in Fig.~\ref{fnlplot}. Once again we find that the expansion method fails in the regions where the power spectrum is most suppressed
and the error bars appear largest.

\subsubsection{Case 3 : $\lambda = 10^{-16}$}

For smaller values of $\lambda$, $\mathcal{P}_{\mathrm {inf}}$ is of course even smaller, and  
it is expected that $\mathcal{P}_{\mathrm {pre}}$ will be too. It is however interesting to 
explore such cases for the following reason. If the regular $\delta N$ formalism was applicable one 
would expect both contributions to $\zeta$ to scale in proportion to 
the energy scale at horizon crossing, but there is no guarantee this will happen in 
the non-perturbative case. 

For $\lambda = 10^{-16}$ one finds  $\mathcal{P}_{\mathrm {inf}} =1.19 \times 10^{-12}$, 
$H_{\mathrm{ini}} = 7.59 \times 10^{-8} M_{\mathrm{pl}}$ and 
$ \langle \delta \chi^2 \rangle = 8.028 \times 10^{-15} M^2_{\mathrm{pl}}$. 
For this case we  
choose representative examples $\bar \chi_i = (i) \times 10^{-7} M_{\mathrm{pl}}$ where $i$ runs from $5$ to $12$,  
and our results are also presented in Fig.~{\ref{Pplot}. Once again $\mathcal{P}_{\mathrm {pre}} $ is smaller 
than $\mathcal{P}_{\mathrm {inf}} $, but we see that in this case it is suppressed by a smaller factor. Again for interest 
we can calculate $\fnl$, and present the results in Fig.~\ref{fnlplot}.

\subsubsection{Case 4 : $\lambda = 10^{-18}$}
Finally we consider $g^2/\lambda = 2$ and  $\lambda = 10^{-18}$. In this case  we have 
$\mathcal{P}_{\mathrm {inf}} = 1.19 \times 10^{-14}$, $H_{\mathrm{ini}} = 7.59 \times 10^{-9} M_{\mathrm{pl}}$ and  $\langle \delta \chi^2 \rangle = 8.028 \times 10^{-17} M^2_{\mathrm{pl}} $ .  We pick $\bar \chi_i = (i) \times 10^{-8} M_{\mathrm{pl}}$ where $i$ runs from $5$ to $12$. Results are shown in Fig.~\ref{Pplot} and \ref{fnlplot} for the amplitude of the power spectrum and interestingly we now find that the preheating contribution, ${\cal P}_{\rm pre}$, 
is greater than the inflationary contribution. 

\subsubsection{Discussion}

It is interesting to note that in all cases the typical value of $\mathcal{P}_{\mathrm {pre}}$ is very similar. This can be understood 
by noting that here we have taken a range of $\bar \chi$ that is scaled in proportion to $1/\sqrt{\lambda}$ in each case, as is the square root of the variance of $\chi$ at the initial time. Moreover for this particular system, we find that the effect of reducing $\lambda$ 
is to shift the pattern of spikes seen in Fig.~\ref{L14ini}} to lower values of $\chi$ roughly in proportion to $\sqrt{\lambda}$, but are otherwise very similar. This is a consequence of the conformal invariance of the system. Since all elements of the calculation that go into evaluating Eq.~(\ref{npdn}) scale in the same way, 
the answer is largely unchanged. This is in stark contrast with the contribution to $\zeta$ from the inflation. Nevertheless, the resulting amplitude ${\cal P}_{\rm pre}$ is smaller 
than the observed value, and therefore this mechanism cannot account for the observed curvature perturbations.

In cases where the expansion method can be trusted, the spectral index of the preheating 
contribution is inherited directly from the 
spectrum of the isocurvature fluctuations at the initial time, but given the lack of compatibility with observations we don't 
pursue its calculation further.

\section{Conclusions}

The main purpose of this work was to show that through lattice simulations it is possible 
to calculate the power spectrum and bispectrum of the primordial curvature 
on observational scales when a preheating field 
plays a significant role. We have demonstrated this explicitly by considering the massless preheating model. 
We found in all the cases we looked at, that even when the homogeneous mode 
is within the strong resonance regime (when $g^2/\lambda\approx 2$),
the contribution from the preheating field to the power spectrum is much lower than the observed value.
We did find however that it could dominate over the inflaton contribution because it remains roughly constant as the energy 
scale is lowered, in contrast to the contribution from the inflation. This is in stark contrast to the expected behaviour 
for a contribution produced during inflation.  We also found that it is highly 
sensitive to the mean value of the reheating field, and that it is roughly 
 an order of magnitude larger than that found in 
earlier analytical studies. These findings indicate the importance of calculating this 
contribution on a case by case and model by model basis. Finally, we also experimented with reducing the ratio
${g^2}/{\lambda}$,  which gradually moves the homogeneous 
mode out of the preheating resonance band, and as expected the contribution of the reheating field 
to the curvature perturbation decreased further in these cases.

For the value of $\lambda$ for which the inflaton contribution to the curvature perturbation
provides the observed amplitude, we found that the preheating field leads to a subdominant contribution to the power spectrum, but can provide the dominant contribution to the bispectrum for some range of 
initial conditions.

It is worth contrasting our study with earlier ones. In contrast to early work \cite{Chambers:2007se,Chambers:2008gu,Bond:2009xx,Chambers:2009ki}, which we otherwise follow closely, 
we do not rely on a Taylor expansion in $\delta N$, but instead compute the statistics of the curvature perturbations on large scales using an expansion in powers of the field correlator.
Our methods and aims are very similar to those of \cite{Suyama:2013dqa}, but that 
work did not work directly with simulation data, but rather the previously generated function in the work of Ref.~\cite{Bond:2009xx}.

Our study relied on an expansion of the full non-perturbative $\delta N$ formalism described in 
Ref.~\cite{Imrith:2018uyk}, and we found that this was  valid except in cases where the leading contribution dropped 
off significantly. It would 
be possible to do better using the fully non-perturbative method as described in that work, but this would require us to run 
many more simulations. Moreover we 
assumed that the inflaton contribution and the preheating field's contribution to $\zeta$ were uncorrelated, 
and we could treat the preheating field as a light Gaussian spectator right up to the point it becomes massive. In general, one 
can do better than these approximations either analytically or by using codes such as that described in Ref.~\cite{Dias:2016rjq}.
In the present case to go beyond what we have done is not warranted, given that it would be very 
unlikely to change the incompatibility of the 
scenario with observations. Nevertheless for more realistic models, for example that of Ref.~\cite{Chambers:2009ki}, one might 
need to turn to these methods to confront models that need to be studied with lattice simulations with observations. The 
present work, however, establishes the feasibility of doing this.

\section*{Acknowledgements} 
DJM is supported by a Royal Society University Research Fellowship and SVI acknowledges the support of the STFC grant ST/M503733/1. AR is supported by the STFC grant ST/P000762/1. 

\bibliography{mybib}

\end{document}